\documentstyle[sprocl,epsfig]{article}
\begin{document}
\title{Overview of Kaon Physics}
\author{R. D. Peccei}
\address{Department of Physics and Astronomy, UCLA, Los Angeles, CA 
90095-1547}
\maketitle
\begin{abstract}

In this overview, I discuss some of the open issues in Kaon physics. After briefly touching on lattice calculations of Kaon dynamics and tests of CPT, my main focus of attention is on $\epsilon^\prime/\epsilon$ and on constraints on the CKM model. The impact of rare K- decays and of experiments with B-mesons for addressing the issue of CP-violation is also discussed. The importance of looking for signals of flavor-conserving CP violating phases is emphasized.

\end{abstract}

\section{Introductory Remarks}

It is difficult to overview a mature field like Kaon 
physics.  In thinking about the subject, it seemed natural
for me to divide my report into three parts.  The first of these
parts addresses areas of Kaon physics where steady progress
continues to be made, but more still needs to be done.  Kaon
dynamics, as well as tests of conservation laws and of the
CKM Model~\cite{CKM} properly belong in this category.  The
second part encompasses what, colloquially, might be called the
``hot topic" of Kaon physics---the new experimental results on
$\epsilon^\prime/\epsilon$ and their theoretical interpretation.
Finally, in the last part, I grouped together topics in Kaon physics
which bring new insights into the future, particularly
regarding the nature of flavor.  Although  CP-violation in B-decays
obviously has little to do directly with Kaons its study, along with
that of rare Kaon decays, properly belongs in this last category.  So
does the hunting for new CP violating phases.

\section{Kaon Decay Dynamics}

By contrasting the ratio of the charged to neutral Kaon
lifetimes [$\tau(K^+)/\tau(K^o_S)$ $ = 138.6 \pm 0.4$~\cite{PDG}]
with that for $B$-mesons [$ \tau(B^+)/\tau(B^o) = 1.072 \pm 0.026$~\cite{PDG}],
it is clear that Kaon decays involve strong interaction
dynamics.  While $B$-decays are essentially reflective of the
decays of the $b$-quark, Kaon decays are strongly dependent
on the underlying QCD dynamics.

A long-standing puzzle of Kaon decays has been the, so
called, $\Delta I=1/2$ rule, which encodes the dominance of the
$I = 0~\pi\pi$ final state in Kaon decays.   Why this should be
so, and what fixes the ratio of isospin amplitudes
\begin{equation}
\frac{A_2(K\to\pi\pi)}{A_0(K\to\pi\pi)} \simeq
\frac{1}{22}
\end{equation}
is not really known.  Although QCD gives a significant short
distance enhancement to the $\Delta I=1/2$ matrix element in
the $K\to 2\pi$ amplitudes,~\cite{AMGL} the
dominant effect appears to be non-perturbative in nature.
So, to estimate theoretically the ratio (1) requires lattice
or $1/N_c$ methods.

Recent lattice QCD results for $A_2/A_0$~\cite{PK}
show a significant enhancement for this ratio, but do not
yet reproduce the experimental result (1). There are several reasons one can adduce for the, roughly, factor of two discrepancy betwen theory and experiment.  First, the results obtained are quite sensitive to 
{\bf chiral corrections}. Technically what is studied are the
matrix elements of operators between a Kaon and pion state
$[\langle\pi|O_i|K\rangle]$, rather than the matrix elements
of the relevant operators between a Kaon and two pions.  
These quantities are related exactly only in the soft pion
limit, so one must carefully correct for this.  In addition, 
in the recent calculation of Pekurovsky and Kilcup,~\cite{PK} it appears
that $A_2$ itself is strongly dependent on the $m^2_K\to 0$
extrapolation performed.  Finally, for accurate results, fully dynamical quarks must be included. However, present results 
 ~\cite{PK} do not appear to show much
difference between quenched and unquenched calculations.  

The inclusion of dynamical quarks, perhaps, is  more critical to the attempts to extract a reliable value for the strange
quark mass-- a quantity which is of importance for  $\epsilon'/\epsilon$.  This is nicely illustrated in Fig. 1, which shows that in
the quenched limit the values of $m_s$ are quite sensitive to the
discretization used.\footnote{These values are also quite sensitive to 
the physical input used.  For instance, the CP-PACS collaboration~\cite{CPP}
reports value for $m_s$ which differs by about 30 MeV, depending 
on whether they used Kaons or Phi mesons as input.}  From
this data one can estimate a quenched value~\cite{ms} for the
strange quark mass of
\begin{equation}
m_s(2~{\rm GeV}) = (100 \pm 20)~{\rm MeV}~.
\end{equation}
My sense is that this value still 
{\bf overestimates} the true value.  However, it is not clear by how much! 

\begin{figure}[t]
\center
\epsfig{file=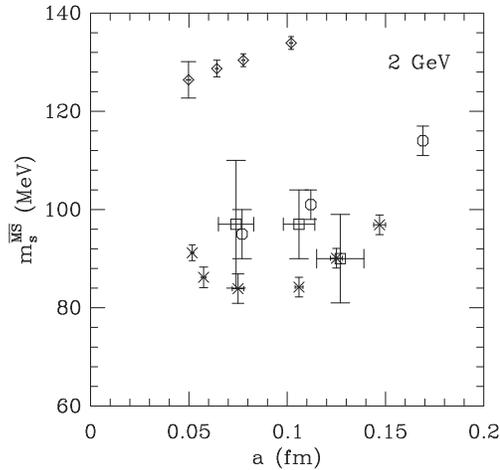,height=3in}
\caption{Results for $m_s$ in the quenched approximation, for different discretizations, from Ref. 6.}
\end{figure}

\section{Testing Conservation Laws}

Almost from their discovery, Kaon decays have been a 
marvelous test-bed for conservation laws, and have helped us
improve our understanding of particle interactions.  The tests
done with Kaons have ranged from tests of fundamental
conservation laws, like CPT, to more mundane tests
like those connected to the $\Delta S=\Delta Q$ rule, which
basically just reflects the interactions allowed by the
Standard Model.  The most famous tests of conservation laws
in the Kaon system, of course, deal with CP-violation.  I will
address this subject in much more detail in the coming Sections.
Here, I concentrate on two tests
that help probe physics at very large scales---scales much
larger than the scale of the weak interactions
$v_F \sim (\sqrt{2}~G_F)^{-1/2} \simeq 250~{\rm GeV}$.

\subsection{Lepton Flavor Violation}

The first of these tests involve lepton flavor violation.
Here strong bounds exist, in both neutral Kaon decays~\cite{BNL871} 
\begin{equation}
{\rm BR}(K_L^o \to \mu e) < 4.7\times 10^{-12}~~~
(90\%~{\rm C.L.})
\end{equation}
and charged Kaon decays~\cite{BNL777}
\begin{equation}
{\rm BR}(K^+ \to \pi^+\mu e) < 2.1\times 10^{-10}~~~
(90\%~{\rm C.L.})~,
\end{equation}
with more refined values expected at this Conference.
These bounds, typically, can be used to pin down limits on the
scale of possible new physics associated with lepton
flavor violation.  For instance, the existence of massive
leptoquarks would violate lepton
number. Assuming that the leptoquark coupling to quarks
and leptons is of electroweak strength, $t$-channel leptoquark exchange   
gives a branching ratio for the
process $K_L\to\mu e$:
\begin{equation}
{\rm BR}(K_L\to\mu e)\sim \left(\frac{M_W}{M_{LQ}}\right)^4~.
\end{equation}
Thus, branching ratios of order $10^{-12}$ probe leptoquark masses in the 100 TeV range.

\subsection{CPT Violation}

The CPT theorem~~\cite{CPT} is based on rather sacred principles.
Any local, Lorentz invariant, quantum field theory with the
normal spin-statistics connection conserves CPT.  Thus, the
experimental observation of a signal that violates CPT would be a
spectacular discovery.  Theoretically, a violation of CPT
can also occur through a violation of quantum mechanics.~\cite{QM}
Corrections to the
Schr\"odinger equation for the density matrix of the form
\begin{equation}
i\frac{\partial}{\partial t}\rho = H\rho - \rho H^\dagger +
\delta h\rho
\end{equation}
will produce phenomena which effectively violate CPT, if $\delta h\not= 0$.

In the following, for simplicity, I will 
assume that quantum mechanics is valid. Even in this case, the phenomenology of CPT-violation is quite
rich.  Essentially, CPT-violation causes two modifications to 
the usual $K^o-\bar K^o$ formalism:~\cite{BCDP}
\begin{description}
\item{i)} The diagonal elements of the effective Hamiltonian
for the system, \break $H = M-\frac{1}{2}\Gamma$, are no longer equal.
This introduces into the formalism a CPT-violating
parameter
\begin{equation}
\delta = \frac{M_{K^o}-M_{\bar K^o} - \frac{i}{2}
(\Gamma_{K^o}-\Gamma_{\bar K^o})}
{2\left[m_S-m_L-\frac{i}{2}(\Gamma_S-\Gamma_L)\right]}~.
\end{equation}
Here $m_{S,L}$ and $\Gamma_{S,L}$ are, respectively, the masses
and width of the physical eigenstates of the $K^o-\bar K^o$
complex, the short- and long-lived $K$-mesons, $K_S$ and
$K_L$.
\item{ii)} The decay amplitudes for any physical processes
are twice as many, since particle and antiparticle decays are
no longer simply connected by charge conjugation (modulo 
strong rescattering phases).  Thus, for example, one has~\cite{Okun}
\begin{equation}
\langle 2\pi;I|T|K^o\rangle =
(A_I+B_I)e^{i\delta_I};~~
\langle 2\pi;I|T|\bar K^o\rangle = (A_I^*-B_I^*)e^{i\delta_I} 
\end{equation}
\begin{equation}
\langle\pi^-\ell^+\nu_\ell|T|K^o\rangle = a+b;~~~
\langle \pi^+\ell^-\bar\nu_\ell|T|\bar K^o\rangle = a^*-b^*~,
\end{equation}
where the amplitudes $B_I$ and $b$ violate CPT.
\end{description}

The principal test of CPT in the $K^o-\bar K^o$ system at the
moment comes from measurements of the parameter $\epsilon$,
connected with the amplitude ratio of the $K_L$ and $K_S$
amplitudes into two pions.  One can show~\cite{BCDP} that
$\epsilon$ has two components, one CP-violating and one
CPT-violating:
\begin{equation}
\epsilon = \epsilon_{C{\not P}} \exp\left[i\phi_{SW}\right] +
 \epsilon_{C{\not P}T} \exp\left[i\left(\phi_{SW} +
\frac{\pi}{2}\right)\right]~.
\end{equation}
Here $\phi_{SW}=\tan^{-1}[2\Delta m/(\Gamma_S-\Gamma_L)]= (43.64\pm 0.08)^\circ~$~\cite{PDG}, while
\begin{equation}
\epsilon_{C{\not P}T} \simeq\sqrt{2} {\rm Im} \delta \simeq\sqrt{2}(\frac{{\rm Re B_o}}{{\rm Re A_o}}- {\rm Re}\delta)~.
\end{equation} 
Because, $\phi_{+-}\simeq \phi_{SW}$, it is clear that $ \epsilon_{C{\not P}}>>
\epsilon_{C{\not P}T}$. Thus, the decay $K_L \to \pi\pi$ is a sign of CP-violation, not CPT-violation. Since, $\epsilon^\prime$ is small, one can infer a value for $\epsilon_{C{\not P}T}$ from the meausurement of $\phi_{+-}$ ~\cite{PDG}. In this way, one arrives at the following values for the CPT violating parameters:
\begin{equation}
 {\rm Im}\delta =(\frac{{\rm Re B_o}}{{\rm Re A_o}}- {\rm Re}\delta)= (0.29 \pm 1.58)\times 10^{-5}.
\end{equation} 

With only this meausurement, it is not possible to bound the $K^o-\bar K^o$ mass difference, since this difference depends on both $ {\rm Re}\delta$ and ${\rm Im}\delta $. However, the CPLEAR collaboration has recently meausured  $ {\rm Re}\delta$ independently,  by studying the asymmetry in the time evolution of semileptonic $K^o$ and $\bar K^o$ decays. They find~\cite{CPLear}
\begin{equation}
 {\rm Re}\delta= (3.0 \pm 3.3 \pm 0.6)\times 10^{-4}.
\end{equation}
This result should be improved by KLOE to perhaps the $10^{-4}$ level. At any rate, one can now bound
\begin{equation}
 |M_{K^o}-M_{\bar K^o}|=2\Delta m({\rm Im}\delta - {\rm Re}\delta)= (2.1\pm 2.8)\times 10^{-18}~.
\end{equation} 
This is an extremely stringent test of CPT indeed!

\section{Testing the CKM Paradigm}

The CKM Model~\cite{CKM} is the simplest example of a model with CP-violation. With three generations of quarks, the mixing matrix $V_{\rm CKM}$ in the charged current has {\bf one} CP violating phase. \footnote{The model has also a CP violating angle $\bar \theta$, which must be extremely tiny to respect the strong bounds one has on the neutron dipole moment. Why this should be so, is not really understood.~\cite{RDP}} More importantly, the model gives a consistent, and  qualitatively understandable, description of the observed phenomena. In particular, the CP violating parameter $\epsilon$ is small not because the CP violating phase $\gamma$ in $V_{\rm CKM}$ is small, but as the result of the smallnes of the intergenerational mixing.

The consistency of the CKM model with the observed CP-violating phenomena
in the Kaon system emerges from a careful study of constraints on the CKM
mixing matrix.  It is useful for these purposes, following Wolfenstein,\cite{Wolf} to expand the elements of $V_{\rm CKM}$ in powers of 
the Cabibbo angle $\lambda = \sin\theta_c = 0.22$:
\begin{eqnarray}
V_{\rm CKM}
&\simeq&
\left(
\begin{array}{ccc}
1-\lambda^2/2 & \lambda & A\lambda^3(\rho-i\eta) \\
-\lambda & 1-\lambda^2/2 & A\lambda^2 \\
A\lambda^3(1-\rho-i\eta) & -A\lambda^2 & 1
\end{array}
\right)
+ O(\lambda^4)~.
\end{eqnarray}
One sees from the above that, to $O(\lambda^4)$, the only complex phases
in $V_{\rm CKM}$ enter in the $V_{ub}$ and $V_{td}$ matrix elements:
\begin{equation}
V_{ub} = A\lambda^3(\rho-i\eta) \equiv |V_{ub}| e^{-i\gamma}; ~~~
V_{td} = A\lambda^3(1-\rho-i\eta) \equiv |V_{td}| e^{-i\beta}~.
\end{equation}
The unitarity condition
$\sum_i V^*_{ib}V_{id} = 0$
on the $V_{\rm CKM}$ matrix elements has a nice geometrical interpretation
in terms of a triangle in the $\rho-\eta$ plane with base $0\leq\rho\leq 1$
and with an apex subtending an angle $\alpha$, where $\alpha + \beta +
\gamma = \pi$.

One can use experimental information on $|\epsilon|$, the $B_d-\bar B_d$
mass difference $\Delta m_d$, and the ratio of $|V_{ub}|/|V_{cb}|$ inferred
from $B$-decays to deduce a 95\% C. L. allowed region in the $\rho-\eta$ plane.
If one includes, additionally, information from the recently obtained
strong bound on $B_s-\bar B_s$ mixing [$\Delta m_s > 12.4~{\rm ps}^{-1} ~~$(95\% C.L.)~\cite{smix}], one further restricts the CKM allowed region.
Fig. 2 shows the result of a recent study for the Babar Physics Book.~\cite{Babar} As one can see, the data is consistent with a rather
large CKM phase $\gamma :~45^\circ \leq \gamma \leq 120^\circ~$.
If one were to imagine that $|\epsilon|$ is due to some other physics, as
in the superweak theory,~\cite{SW} then effectively the $\Delta S = 1$
parameter $\eta \simeq \gamma \simeq 0$. In this case one has another allowed 
region for $\rho$ at the 95\% C.L.: $ 0.25 \leq \rho \leq 0.27$~.

\begin{figure}[t]
\begin{center}
\epsfig{file=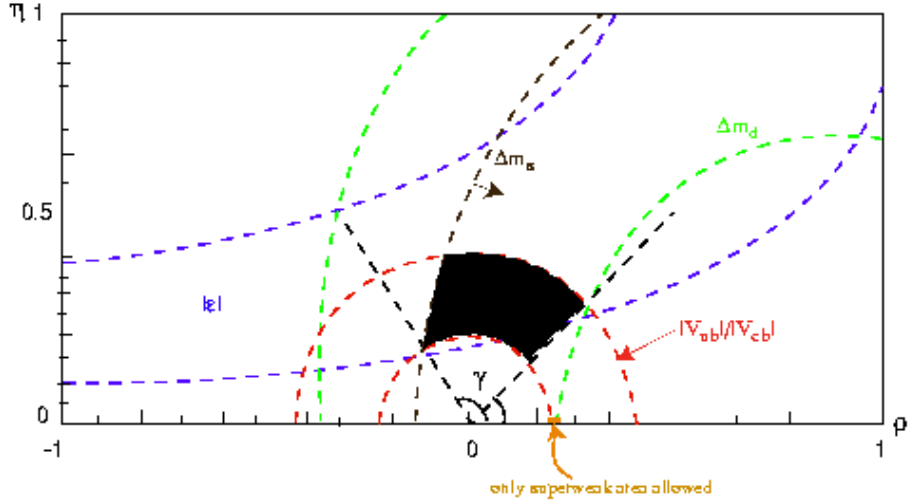,width=5in}
\end{center}
\caption[]{Allowed region in the $\rho-\eta$ plane. From Ref. [16]}
\end{figure}

Two experimental results in 1999 have strengthened the case for the validity of the CKM model:
\begin{description}
\item i) The CDF collaboration~\cite {CDF} has announced a first significant result for $\sin 2\beta$ from a study of $B \to \Psi K_S$ decays
\begin{equation}
\sin 2\beta=0.79^{\scriptstyle +0.41}_{\scriptstyle -0.44},
\end{equation}
whose central value coincides with that emerging from the fit of Fig. 2
\item ii)  Recently, 
the KTeV collaboration announced a new result for $\epsilon^\prime/\epsilon$
obtained from an analysis of about 20\% of the data collected in the last two
years.~\cite{KTeV}  Their result 
\begin{equation}
{\rm Re}~\epsilon^\prime/\epsilon = (28.0 \pm 4.1)\times 10^{-4},
\end{equation}
is much closer to the old CERN result [${\rm Re}~\epsilon^\prime/\epsilon =
(23 \pm 6.5) \times 10^{-4}$],~\cite{NA31} 
than  to the value obtained by KTeV's precursor[${\rm Re}~\epsilon^\prime/\epsilon = (7.4 \pm 5.9) \times 10^{-4}$].~\cite{E731}  More significantely, the value obtained is nearly $7\sigma$ away from zero, giving strong evidence that the CKM phase $\gamma$ is indeed non-vanishing.
\end{description}

\section{Lessons from $\epsilon^\prime/\epsilon$}

If one were to average the KTeV,~\cite{KTeV} NA31~\cite{NA31} and E731~\cite{E731} results for $\epsilon^\prime/\epsilon$, the resulting number[ ${\rm Re}~\epsilon^\prime/\epsilon = (21.8 \pm 3.0) \times 10^{-4}$] is in marvelous  agreement with the result announced at Kaon99 by the NA48 Collaboration:~\cite{Sozzi}
\begin{equation}
{\rm Re}~\epsilon^\prime/\epsilon = (18.5 \pm 4.5 \pm 5.8)\times 10^{-4}.
\end{equation}
This is very strong evidence that there exists $\Delta S=1$ CP-violation, which is the chief premise of the CKM Model.

A non zero  $\epsilon^\prime/\epsilon$ is good news for the CKM model. However, a value for this ratio of around $20 \times 10^{-4}$ is unsettling, since this result appears a bit too large. This can be seen from Table 1, which  displays some recent  theoretical expectations for  $\epsilon^\prime/\epsilon$.
What is particularly perturbing is that the smaller values in the Table correspond to, in principle, more theoretically pristine calculations of the relevant matrix elements, based on lattice methods. However, the calculation of 
$\epsilon^\prime/\epsilon$ is challenging, since large cancellations are involved.

\begin{table}
\caption{Recent theoretical results on $\epsilon^\prime/\epsilon$}
{\begin{center}
\begin{tabular}{|c|c|} \hline
$\epsilon^\prime/\epsilon \times 10^4$ & Reference\\  \hline
$4.6 \pm 3.0 \pm 0.4$ & Ciuchini {\it et al.}~\cite{Ciuc} \\
$5.2^{\scriptstyle +4.6}_{\scriptstyle -3.7}$& Bosch {\it et al.}~\cite{Bosc}\\
$7.7^{\scriptstyle +6.0}_{\scriptstyle -3.5}$& Bosch {\it et al.}~\cite{Bosc}\\
5--22& Paschos {\it et al.}~\cite{Pas}\\
$17^{\scriptstyle +14}_{\scriptstyle -10}$& Bertolini {\it et al.}~\cite{Bert}\\ \hline
\end{tabular}
\end{center}}
\end{table}

As is well known, $\epsilon^\prime/\epsilon$ depends on both the matrix elements of gluonic and electroweak Penguin operators.~\cite{BBL} The former are enhanced over the latter, since they are of $O(\alpha_s)$ rather than of $O(\alpha)$. However, the electroweak Penguins are not suppressed by the $\Delta I=1/2$ rule and are enhanced by the large top mass.~\cite{FR} As a result, these two contributions are comparable in size. What makes matters worse, is that these
contributions tend to cancel each other, increasing the uncertainty in the
theoretical predictions. One can appreciate the nature of the problem from an approximate formula for $\epsilon^\prime/\epsilon$ due to Buras and collaborators:~\cite{Bosc}
\begin{equation}
{\rm Re}~\epsilon^\prime/\epsilon \simeq
34\eta\left[B_6 - 0.53 B_8\right]\left[\frac{110{\rm MeV}}{m_s({\rm 2 GeV})}\right]^2\times 10^{-4}~.
\end{equation}
In the vacuum insertion approximation, the contribution of the gluonic Penguins, $B_6$, and that of the electroweak Penguins, $B_8$, are both equal to one. Since $\eta\simeq 0.3$, in this approximation, one expects $\epsilon^\prime/\epsilon \simeq 5 \times 10^{-4}$. 

To get agreement with experiment, one needs an appropriate linear combination of four things to happen: $\eta$ should be maximized [$\eta\simeq 0.4$ ?]; the strange quark mass should be smaller [$ m_s({\rm 2 GeV})\simeq 90{\rm Mev}$ ?]; $B_6$ should be bigger than unity [$B_6\simeq 2$ ?]; $B_8$ should be smaller than unity [$B_6\simeq 0.5$ ?]. As Bosch
{\it et al.} remark in their recent analysis,~\cite{Bosc} it is possible to stretch the parameters to get $\epsilon^\prime/\epsilon \simeq 20\times 10^{-4}$, but it is not easy. What is agreed is that the gluonic Penguin matrix elements are very uncertain in the lattice. However, whether $B_6$ can get much bigger than one, as it appears to be in the chiral quark model~\cite{Bert}, remains to be seen. On the other hand, both lattice and 1/N estimates~\cite{Pas} suggest that $B_8$ can be  quite a bit smaller than unity.

The apparent discrepancy between theory and experiment for $\epsilon^\prime/\epsilon$ has spurred a number of people to invoke new physics explanations. For instance, if somehow the $Zds$ vertex were anomalous,~\cite{Zds} then one could get a bigger value for $\epsilon^\prime/\epsilon$. \footnote{One needs $Im Z_{ds}\simeq -(2.5-8) \times 10^{-4}$, rather than the SM value of around $ +1\times 10^{-4}$ for this parameter.}. However, it is not clear what the physical origin of such an anomaly is. Chanowitz~\cite{Chan}  has tried to relate this anomaly to that which seems to affect $Z \to b\bar b$. However, it turns out that this connection gives the wrong sign for $\epsilon^\prime/\epsilon$! Nevertheless, if such an anomalous $Zds$ vertex existed, it would substantially increase the branching ratio for $K_L \to \pi^o \nu \bar\nu$.~\cite{CI} Masiero and Murayama,~\cite{MM} on the other hand, noticed that in supersymmetric extensions of the SM it is possible to get contributions to $\epsilon^\prime/\epsilon$ of order $10^{-3}$ from rather natural $\tilde s_R \tilde d_R$ squark mixings. Thus they suggested that, perhaps,  the large value of $\epsilon^\prime/\epsilon$  is the first  experimental manifestation of low energy supersymmetry at work.

I am personally quite skeptical that the $\epsilon^\prime/\epsilon$ result is a signal of new physics. In my view, the most likely explanation for the discrepancy between theory and experiment is rooted in our inability to accurately calculate K-decay matrix elements. Our long and frustrating experience with the $\Delta I=1/2$ rule should provide an object lesson here! Perhaps the most naive conclusion to draw is that, no matter what else is causing $\epsilon^\prime/\epsilon$ to be large, the CKM parameter $\eta\sim \sin \gamma$ lies near its maximun. That is, perhaps $\gamma \simeq \pi/2$. I do not know a particular reason why this should be so.\footnote{Note that this is not the same as the idea of maximal CP violation, suggested by Fritzsch and others.~\cite{Fritzsch}} However, if this is so, there is an interesting phenomenologigal consequence. It is easy to see that  $\gamma \simeq \pi/2$ predicts that
\begin{equation}
\sin 2\alpha\simeq\sin2\beta\simeq \frac{2\eta}{1+\eta^2}~.
\end{equation}
This "prediction" is at the edge of the CKM fits, but suggest that $\sin 2\alpha$, like $\sin 2\beta$, is also large.

\section{Grappling with the Unitarity Triangle}

From the above discussion, it is clear that to make progress one will
need further experimental input.  Fortunately, help is on the way!

\subsection{CP-Violation in B-Decays}

With the turn-on of the B-factories at SLAC and KEK, and with the upgrade of the
Tevatron with the Main Injector, a new experimental era in the study of CP-violation is beginning. Data which will be collected by the new Babar and Belle detectors, and with CDF and DO, should permit testing the unitarity 
triangle through separate measurements of $\alpha$, $\beta$ and
$\gamma$.  As I alluded to earlier, one of the most robust predictions of the
CKM model is that $\sin 2\beta$ should be large. In contrast to $\epsilon^\prime/\epsilon$, this parameter can be extracted in a
theoretically clean way by studying B-decays to CP self-conjugate
states.~\cite{BKUS}  The decay probability of a state which at
$t=0$ was a $B_d$ into a CP-self conjugate final state, like $\Psi K_S$, has a time evolution which directly isolates $\sin 2\beta$:
\begin{equation}
\Gamma(B_d^{\rm Phys.}(t)\to \Psi K_S) \sim
e^{-\Gamma_Bt}\left[1 + \sin 2\beta\sin\Delta m_dt\right]~.
\end{equation}
To measure  $\sin 2\beta$ experimentally one needs to be able to tag the initial  state as a $B_d$ and then to follow its time development. Once this is achieved,  as was done recently by CDF, ~\cite{CDF} then Eq. (22) yields
$\sin 2\beta$ with essentially no theoretical error.

A study done in preparation for the turn-on of the SLAC B factory~\cite{Babar} estimates that, with an integrated luminosity of $30 {\rm fb^{-1}}$, one could meausure $\sin 2\beta$ with 
an error of the order of $\delta\sin 2\beta \leq 0.08$ .  However, to really test the CKM model, 
measuring $\sin 2\beta$ is not enough. What one wants really is to
measure  also the  other two angles in the unitarity triangle, to see if the triangle indeed closes. In addition,
getting a clean measurement of the Wolfenstein
parameter $\eta$ would be helpful, as this parameter meausures the height of the triangle and hence provides redundant information.  However, extracting $\alpha$ and $\gamma$ from B-decays to comparable accuracy to that with which $\beta$ will be known is likely to be challenging. To reduce the error on these quantities a variety of processes will need to be studied.

Let me briefly illustrate the nature of the problem by discussing how to obtain $\alpha$ in a manner analogous to $\beta$. It is easy to see, ~\cite{BKUS} {\it mutatis mutandi}, that the time development of the decay $B_d^{\rm phys}\to\pi^+\pi^-$, provides information on $\alpha$.  However, for the decay $B_d\to \pi^+\pi^-$, the quark decay amplitude $b\to u\bar ud$ and the Penguin amplitude entering in the process depend differently on the weak  CP-violating phases. The quark decay is proportional to $e^{-i\gamma}$, while the Penguin piece is proportional to $e^{-i\beta}$. \footnote{  There is no Penguin pollution~\cite{penguin} for
$B_d \to \Psi K_S$, since in this case both the quark decay amplitudes and the Penguin amplitudes have the same weak phase.} Only by neglecting the Penguin contributions altogether, one arrives in this case to a formula like (22), with $\beta \to \alpha$.

It is possible to estimate and correct for the Penguin pollution through an isospin analysis of various 
channels.~\cite{Gronau} This, however, exacts a price in precision. Although only time and real data will tell, it
appears difficult for me to imagine measuring $\sin 2\alpha$ to better than $\delta\sin 2\alpha = 0.15$.  Coupled with estimates of the error with which one can extract the CP-violating phase $\gamma$ from B-decays,\cite{Fleischer} my guess is that probably one will not be able to bring down the error on the sum of the angles in the unitarity triangle to better than $\delta[\alpha + \beta + 
\gamma] \simeq (20-30)^\circ$.

\subsection{$K_L\to \pi^o\nu\bar\nu$}

In this respect, the process $K_L\to\pi^o\nu\bar\nu$ offers an interesting alternative opportunity to test the CKM model.  This decay allows a theoretically very clean extraction of $\eta$, and hence of the CKM phase $\gamma$. Splitting $K_L$ into its CP-even and  CP-odd parts, the 
amplitude for this process can be written as:
\begin{equation}
A(K_L\to\pi^o\nu\bar\nu) = \epsilon A(K_1\to\pi^o\nu\bar\nu) +
A(K_2\to\pi^o\nu\bar\nu)~.
\end{equation}
However, because the semileptonic decay of the CP-even state $K_1$ is
small and this amplitude is suppressed by a factor of $\epsilon$ in the above, the decay \break $K_L\to\pi^o\nu\bar\nu$ directly meausures  $\Delta S=1$ CP violation.~\cite{Litt}  In addition, the amplitude $K_2\to\pi\nu\bar\nu$ is essentially free of hadronic uncertainties since it depends only on a K to $\pi$ matrix element.  

The NLO QCD analysis of Buchalla and Buras ~\cite{BB} gives
the  following approximate formula, good to 1-2\%, for the  $K_L\to\pi^o\nu\bar\nu$ branching ratio:
\begin{equation}
{\rm BR}[K_L\to\pi^o\nu\bar\nu] = 4.34\times 10^{-4} A^4\eta^2~.
\end{equation}
Using the results of the CKM analysis, leads to the expectation ~\cite{Bosc}
$1.6\times 10^{-11}\leq \Delta R\left[K_L\to\pi^o\nu\bar\nu\right] <
3.9\times 10^{-11}$. However, this process
presents a formidable experimental challenge. Not only is its branching very low , but  one is dealing with an all neutral final state.
Nevertheless, it is clear that an experiment which could probe the process $K_L\to\pi^o\nu\bar\nu$
to a branching ratio of the order of $3\times 10^{-11}$, with an accuracy
better than 25\%, would have a significant impact on the CKM model.  Of course,
as I remarked earlier, if there were indeed an anomalous $Zds$ vertex, then
one may expect branching ratios almost an order of magnitude higher!~\cite{CI}

\section{Looking for Other CP Violating Phases}

It is quite possible that the CKM phase $\gamma$ is the dominant phase
connected with {\bf flavor changing} CP-violation.  Nevertheless, in my opinion,
it is also quite likely that, in addition, there are also some other {\bf flavor conserving} CP violating
phases. The argument is simple. Given that CP is not conserved, renormalizability requires that any interactions in the Lagrangian of
the theory that can be complex should be so.  This means that any interactions beyond the SM involving new fields necessarily always will involve new phases.  For instance,
with two Higgs doublets one has a phase in the mass term connecting these
two fields
\begin{equation}
{\cal{L}}_{H_1H_2} = \mu^2H_1H_2^\dagger + \mu^{2^*}H_1^\dagger H_2~,
\end{equation}
which just cannot be avoided if CP is violated. \footnote{There are other arguments which point towards the existence of other CP violating phases, besides $\gamma$. For instance, one needs to have some non-GIM suppressed CP violating phase in the theory to generate 
the matter-antimatter asymmetry in the Universe 
at the electroweak phase transition.~\cite{Shap}} 

Kaon decays afford a wonderful opportunity to search for the presence of these
flavor conserving CP violating phses.  The nicest example, perhaps, is
provided by quantities where CKM effects either vanish, or are
negligible.  A good case in point is the triple correlation in
$K_{\mu 3}$ decays which meausures the polarization of the outgoing
muon perpendicular to the plane of production, $\langle P_\perp^\mu\rangle$-- a quantity  which vanishes in the CKM model.~\cite{Leurer}

Since $\langle P_\perp^\mu\rangle\sim\langle\vec s_\mu\cdot(\vec p_\mu\times
\vec p_\pi)\rangle$, the transverse muon polarization is T-odd, and therefore is  sensitive to CP violating contributions. However, as Sakurai~\cite{Sakurai} pointed out long ago,  $\langle P_\perp^\mu\rangle$ can also arise
from final state rescattering effects. Fortunately,
although $\langle P_\perp^\mu\rangle$ is affected by final state interactions
(FSI), for the process $K^+\to\pi^o\mu^+\nu_\mu$,  because the final hadron is neutral, these FSI are very small
$[\langle P_\perp\rangle^\mu_{\rm FSI}\sim 10^{-6}$].\cite{Z}
 
The transverse muon polarization is particularly sensitive to any scalar
interactions present in the decay amplitude.
Writing the effective amplitude for the process $K^+\to\pi^o\mu^+\nu_\mu$ as
\begin{eqnarray}
A(K^+\to\pi^o\mu^+\nu_\mu) &=& G_F\lambda f_+(q^2)[p_\alpha\bar u_\mu
\gamma^\alpha(1-\gamma_5)u_{\nu_\mu} \nonumber \\
& & \mbox{}+f_s(q^2)m_\mu\bar u_\mu(1-\gamma_5) u_{\nu_\mu}]~.
\end{eqnarray}
one finds that the transverse muon polarization $\langle P_\perp^\mu\rangle$ is determined by the imaginary  part of the scalar form factor:
\begin{equation}
\langle P_\perp^\mu\rangle \simeq 0.2~{\rm Im}~f_s~,
\end{equation}
with the numerical constant being essentially a kinematical factor.  

The simplest models which have a non trivial ${\rm Im}~f_s$ are
multi Higgs models.~\cite{Swein} In these models one can, in fact, obtain values for
Im $f_s$ which are at the verge of observability. ~\cite{Garisto} Bounds  on Im $f_s$ coming from other
processes allow $\langle P_\perp^\mu\rangle\leq 10^{-2}$ ~\cite{Grossman}  This is precisely the level of sensitivity of a meausurement of 
$\langle P_\perp^\mu\rangle$  done at  Brookhaven 15 years ago:~\cite{Blatt}
\begin{equation}
\langle P_\perp^\mu\rangle = [-3.1\pm 5.3] \times 10^{-3}~.
\end{equation}
An ongoing experiment at KEK, KEK 246, should be able to improve this measurement slightly.  However, it would be very interesting  if one were able to mount an experiment to get to
$\delta\langle P_\perp^\mu\rangle \sim 10^{-4}$.

\section{Concluding Remarks}

We are at a very exciting juncture in Kaon Physics
and in the study of CP-violation. At long last, we can now say that
$\epsilon^\prime/\epsilon$ is at hand. Eventhough, we are now entering an era where crucial information on CP-violation will be learned from B-decays, 
experiments with Kaons will continue to play a central role.  Experiments at the Frascati $\Phi$-factory should further refine our knowledge of fundamental conservation laws, like CPT, and rare Kaon decays will provide precious windows into new phenomena. It is particularly important that, experimentally, we not be afraid to push into obscure corners-- like flavor conserving CP-violation.

On the theoretical front, dynamical calculations of weak decay matrix elements are approaching the level of accuracy needed to really test the theory. However, it is crucial to properly estimate theoretical uncertainties, so that one can better gauge if discrepancies with experiment really signal new physics effects. Indeed, the real issue with $\epsilon^\prime/\epsilon$ is
whether the large value seen represents new physics or old matrix
elements!  

\section*{Acknowledgements}

This work was supported in part by the Department of Energy under contract No. DE-FG03-91ER40662, Task C.


\section*{References}

\begin{thebibliography}{99}

\bibitem{CKM} N. Cabibbo,  Phys. Rev. Lett. 
{\bf 12} (1963) 531; M. Kobayashi and T. 
Maskawa, Prog. Theor. Phys. {\bf 49} (1973) 652.

\bibitem{PDG} Particle Data Group, C. Caso {\it et al.}, Europ. Phys. J. {\bf C3} (1998) 1.
 
\bibitem{AMGL} M. K. Gaillard and B. W. Lee, Phys. Rev. Lett. {\bf 33} (1974) 108; G. Altarelli and L. Maiani, Phys. Lett. {\bf B52} (1974) 351.
 
\bibitem{PK} For a recent discussion, see for example, D. Pekurovsky and  G. Kilcup, hep-lat/9903025, in  Proceedings of the DPF99 Conference, Los Angeles, eds. K. Arisaka and Z. Bern, http://www.dpf99.library.ucla.edu.
 
\bibitem{CPP} CP-PACS collaboration, R. Burkhalter,  in Proceedings of Lattice 98, Nucl. Phys. B. (Proc. Suppl.) {\bf 73} (1999) 3.
 
\bibitem{ms} T. Blum, A. Soni and  M. Wingate, hep-lat/9902016. 
 
\bibitem{BNL871} BNL 871 collaboration, D. Ambrose, {\it et al.} Phys. Rev. Lett. {\bf 81} (1998) 5734.
 
\bibitem{BNL777} BNL 777 collaboration, A. M. Lee {\it et al.}, Phys. Rev. Lett {\bf 64} (1990) 165.
 
\bibitem{CPT} W. Pauli, in {\bf Niels Bohr and the Development of Physics},
ed. W. Pauli (Pergamon Press, New York 1955); J. Schwinger,  Phys. Rev. {\bf 82} (1951) 914; G. L\"uders, 
Dansk Mat. Fys. Medd {\bf 28} (1954) 5; G.
L\"uders and B. Zumino, Phys. Rev. {\bf 110} (1958) 1450.

\bibitem{QM} J. Ellis, J. S. Hagelin, D. V. Nanopoulos and M. Srednicki, Nucl. Phys. {\bf B241} (1984) 381.
 
\bibitem{BCDP} C. D. Buchanan, R. Cousins, C. O. Dib, R. D. Peccei and J. Quackenbush,  Phys. Rev. {\bf D45} (1992) 4088; C. O. Dib and R. D. Peccei, 
Phys. Rev. {\bf D46} (1992)  2265.

\bibitem{Okun} V. V. Barmin {\it et al.}, Nucl. Phys. {\bf B247}(1984) 293: Erratum {\bf B254} (1985) 747; see also, N. W. Tanner and R. H. Dalitz, Ann. Phys {\bf 171} (1986) 463.

\bibitem{CPLear} CPLEAR collaboration, A. Angelopoulos {\it et al.}, Phys. Lett. {\bf B444} (1998) 52.

\bibitem{RDP} For a discussion see, for example, R.D. Peccei,  Nucl. Phys. B (Proc. Suppl.), {\bf 72} (1999) 3. 

\bibitem{Wolf}  L. Wolfenstein, Phys. Rev. Lett. {\bf 51} (1983) 1945.

\bibitem{smix} F. Parodi,  in the  Proceedings of the 29th Conference on High Energy Physics, ICHEP98, Vancouver, Canada, July 1998, eds. A. Astbury, D. Axen and J. Robinson (World Scientific,  Singapore 1999) p1148.

\bibitem{Babar} {\bf The Babar Physics Book}, eds. P. F. Harrison and H. R. Quinn, SLAC-R 504, October 1998.

\bibitem{SW} L. Wolfenstein, Phys. Rev. Lett. {\bf 13} (1964) 562.

\bibitem{CDF} J. Kroll, these Proceedings; see also, CDF collaboration, CDF/PUB/Bottom/CDF/4855.

\bibitem{KTeV} KTeV collaboration,  A. Alavi-Harati {\it et al.}, Phys. Rev. Lett. {\bf 83} (1999) 22.

\bibitem{NA31} NA31 collaboration, G. D. Barr {\it et al.}, Phys. Lett.
B{\bf 317} (1993) 1233.

\bibitem{E731} E731 collaboration, L. K. Gibbons {\it et al.}, Phys.
Rev. Lett. {\bf 70} (1993) 1203.

\bibitem{Sozzi} M. S. Sozzi, these Proceedings

\bibitem{Ciuc}  M. Ciuchini, E. Franco, G. Martinelli and L. Reina,
Phys. Lett. B{\bf 301} (1993) 263; M. Ciuchini, E. Franco, G. Martinelli,  L. Reina and L. Silvestrini, Z. Phys. {\bf C68} (1995) 239. 

\bibitem{Bosc}  S. Bosch, A.J. Buras, M. Gorbahn, S. J\"ager, M. Jamin, M.E. Lautenbacher and L. Silvestrini, hep-ph/9904408.

\bibitem{Pas} E.A. Paschos, DO-TH 99/04, to be published in the Proceedings of the 17th International Workshop on Weak Interactions and Neutrinos (WIN99), Cape Town, South Africa, 1999;  T. Hambye, G.O. Koehler, E.A. Paschos and P.H. Soldan, hep-ph/9906434.
 
\bibitem{Bert} S. Bertolini, J.O. Eeg and M. Fabbrichesi,  hep-ph/9802405, to appear in  Rev. Mod. Phys. ; S. Bertolini, J.O. Eeg, M. Fabbrichesi and E. I Lashin, Nucl. Phys. {\bf B514} (1998) 93.

\bibitem{BBL} G. Buchalla, A. J. Buras and M. E. Lautenbacher, Rev. Mod. Phys. {\bf 68} (1996) 1125.

\bibitem{FR}J. Flynn and L. Randall, Phys. Lett. {\bf B216} (1989) 221;
{\it ibid.} {\bf B224} (1989) 221; Nucl. Phys. {\bf B326} (1989) 3.

\bibitem{Zds}  Y.-Y. Keum, U. Nierste and  A.I. Sanda, Phys. Lett. {\bf B457} (1999) 157; L. Silvestrini, hep-ph/9906202, to appear in the Proceedings of 
the XXXIV Rencontres de Moriond "Electroweak interactions and unified
theories", Les Arcs, France, 1999; see also A. J. Buras and L. Silvestrini,
Nucl. Phys. {\bf B546} (1999) 299.

\bibitem{Chan} M. Chanowitz, hep-ph/9905478.

\bibitem{CI} G. Colangelo and G. Isidori, JHEP {\bf 09} (1998) 009.

\bibitem{MM} A. Masiero and H. Murayama,  Phys. Rev. Lett. {\bf 83} (1999) 907.

\bibitem{Fritzsch}  For a review see, for example, H. Fritzsch, in  in {\bf Broken Symmetries}, eds. L. Mathelitsch and W. Plessas, Lecture Notes in Physics 521 (Springer Verlag, Berlin 1999); see also, A. Mondragon and  E. Rodriguez-Jauregui, hep-ph/9906429.

\bibitem{BKUS} See, for example, I. I. Bigi, V. A. Khoze, N. G. Uraltsev and A. Sanda, in {\bf CP Violation} ed C. Jarlskog (World Scientific, Singapore 1989) p175.
 
\bibitem{penguin} M. Gronau, Phys. Rev. Lett. {\bf63} (1989) 1451; D. London and R. D. Peccei, Phys. Lett. {\bf B223} (1989) 257.

\bibitem{Gronau} M. Gronau and D. London, Phys. Rev. Lett. {\bf 65} (1990) 3381.
 
\bibitem{Fleischer} R. Fleischer, Phys. Lett. {\bf B459} (1999) 306.

\bibitem{Litt} L. Littenberg, Phys. Rev. {\bf D39} (1989) 3322.

\bibitem{BB} G. Buchalla and A. J Buras, Nucl. Phys. {\bf B400} (1993) 225;  {\bf B548} (1999) 309. 

\bibitem{Shap} M. E. Shaposhnikov, Nucl. Phys. {\bf B287} (1987) 757.

\bibitem{Leurer} M. Leurer, Phys. Rev. Lett. {\bf 62} (1989) 1967.

\bibitem{Sakurai} J. J. Sakurai, Phys. Rev. {\bf 109} (1958) 980.

\bibitem{Z} A. R. Zhitnitski, Sov. J. Nucl. Phys. {\bf 31} (1980)
529.

\bibitem{Swein} S. Weinberg, Phys. Rev. Lett. {\bf 37} (1976) 657.

\bibitem{Garisto} R. Garisto and G. Kane, Phys. Rev. D{\bf 44} (1991) 2038;
G. Belanger and C. O. Geng, Phys. Rev. D{\bf 44} (1991) 2789.

\bibitem{Grossman} Y. Grossman, Nucl. Phys. B{\bf 426} (1994) 355.

\bibitem{Blatt} S. R. Blatt {\it et al.}, Phys. Rev. D{\bf 27} (1983) 1056.

\end{thebibliography}
\end{document}